\documentstyle[12pt,aps]{revtex}
\newcommand{\beq}{\begin{equation}}
\newcommand{\eeq}{\end{equation}}
\newcommand{\beqar}{\begin{eqnarray}}
\newcommand{\eeqar}{\end{eqnarray}}
\newcommand{\sect}[2]
{\vspace*{0.5\baselineskip}\hspace*{-\parindent}{\bf #1.}~{\bf
#2}\hspace*{\parindent}}

\begin{document}
\draft

\title{The Off-Shell Axial Anomaly via the
\mbox{\boldmath $\gamma ^* \pi ^0 \rightarrow \gamma $} Transition}
\author{M. R. Frank\footnotemark[1], K. L. Mitchell\footnotemark[2],
C. D. Roberts\footnotemark[3] and P. C. Tandy\footnotemark[2] }
\address{
\footnotemark[1]
Institute for Nuclear Theory, University of Washington, Seattle, Washington
 98195, USA\\
\footnotemark[2]
Center for Nuclear Research, Department of Physics, Kent State University,
Kent, Ohio 44242, USA\\
\footnotemark[3]
Physics Division, Argonne National Laboratory, Argonne, Illinois 60439, USA}
\maketitle
\begin{abstract}
The $\gamma^* \pi^0 \rightarrow \gamma$ form factor, including the
extension off the pion mass-shell, is obtained from a generalized impulse
approximation within a QCD-based model field theory known to provide an
excellent description of the pion charge form factor.
This approach implements
dressing of the vertex functions and propagators consistent with dynamical
chiral symmetry breaking, gauge invariance, quark confinement and
perturbative QCD.
Soft nonperturbative behavior, dictated by the axial anomaly, is found to
evolve to the perturbative QCD limit only for
\mbox{$Q^2 \geq 20~{\rm GeV}^2$}.
\end{abstract}

\sect{1}{Introduction.}
The pion charge form factor for space-like momenta has long been used
as one of the simplest but non-trivial testing grounds for applications
of QCD to hadronic properties.  A closely related quantity that has
received less attention is the $\gamma ^*\pi ^0
\rightarrow \gamma $ transition form
factor.  Here the photon momentum dependence maps out a particular off-shell
extension of the axial anomaly.~\cite{aa}
Presently available data for this transition form factor in the
space-like region $Q^2< 2.5$ GeV$^2$ is from the CELLO~\cite{CELLO}
collaboration at the PETRA storage ring where the process
$e^+e^-\rightarrow e^+e^-\pi ^0$ was measured with geometry requiring
one of the two intermediate
photons to be almost real.  There is currently renewed
interest in this transition form factor of the pion due to the
prospect of obtaining higher precision data over a broader momentum range
via virtual Compton scattering from a proton target at CEBAF.~\cite{cebaf}
In this case a (virtual) pion is supplied by the target and a final real
photon selected through the excellent missing mass spectrometry
available at CEBAF.  It is anticipated that suitable electron scattering
geometry can minimize the t-channel momentum to the extent that the pion
mechanism will dominate other contributions such as
s-channel resonances.~\cite{cebaf}  An extrapolation to the pion mass shell
will be needed to deduce the physical transition form factor.

In this work we present results of a calculation of the $\gamma ^*\pi ^0
\rightarrow \gamma $ vertex as a function of both the virtual photon
and pion momenta.  The mechanism employed is the generalized impulse
approximation or quark triangle diagram in which the confining quark
propagator and the photon-quark and pion-quark vertices are
dynamically dressed quantities consistent with nonperturbative
Schwinger-Dyson equation (SDE) studies~\cite{RW}, and hence, also with
asymptotic QCD.  The constraints of electromagnetic
gauge invariance and dynamical chiral symmetry breaking are obeyed and lead to
the calculation being completely determined by the amplitudes describing the
dynamical quark propagator.  For the latter we take a recently
developed model which has been shown to provide an excellent description
of the space-like pion charge form factor, as well as a variety of soft
chiral-physics quantities.~\cite{pionff}   No adjustment of
parameters is made in the
present application.  The approach naturally identifies the momentum scale
where soft nonperturbative behavior evolves to the hard perturbative QCD limit.

\sect{2}{The \mbox{\boldmath $\pi ^0 \gamma \gamma $} Vertex Function.}
We use a Euclidean-space formulation with metric $\delta _{\mu \nu }$.
The action for the three-point interaction can be written as
\beq
S[\pi ^0\gamma \gamma ]=\int \frac{d^4Pd^4Q}{(2\pi )^8} A_{\mu }(-P-Q)
A_{\nu }(Q)\pi ^0(P)\Lambda _{\mu \nu}(P,Q),\label{action}
\eeq
where $A_\nu$ is the electromagnetic field and the momentum
assignments are shown in Fig.~1.
The general form of the vertex allowed by CPT is
\beq
\Lambda _{\mu \nu }(P,Q)=-i\frac{\alpha }{\pi f_{\pi }}\epsilon
_{\mu \nu \alpha \beta }P_{\alpha }Q_{\beta }~g(Q^2,P^2,P\cdot Q)
\label{vertex}
\eeq
where $\epsilon_{4123}=1$,
$\alpha $ is the fine-structure constant, $f_{\pi }$ is the pion
decay constant, and $g$ is the off-mass-shell invariant amplitude.  With
the one photon mass-shell condition $(P+Q)^2=0$, the invariant amplitude,
denoted by $g(Q^2,P^2)$, is the object of the present work.
For a physical pion the shape of
the $\gamma ^*\pi ^0\rightarrow \gamma $ transition form factor is given
by $g(Q^2,-m^2_{\pi })$.  At the fully on-mass-shell point $(Q^2=0,P^2=
-m^2_{\pi })$, the amplitude for the $\pi ^0\rightarrow \gamma \gamma $ decay
is provided by $2\Lambda _{\mu \nu } \epsilon_\mu(1) \epsilon_\nu(2)$ where
$\epsilon_\mu$ is a photon polarization vector.  The chiral limit for the
strength of this decay amplitude is fixed at $\frac{\alpha }{\pi f_{\pi }}$
by the axial anomaly~\cite{iz} which gives an
excellent account of the $\pi ^0\rightarrow \gamma \gamma $ decay width of
$7.7$ eV.  Thus $g(0,0)=1/2$ follows only from gauge invariance and chiral
symmetry in quantum field theory and provides a stringent check upon model
calculations.
\begin{figure}
\unitlength1.cm
\begin{picture}(14,8)(-12.0,-3.5)
\includegraphics{ggpi.ps}
\end{picture}
\caption{The quark triangle diagram for the generalized impulse approximation
to the $\gamma^* \pi^{0} \gamma$
vertex.}
\label{fig1}
\end{figure}

Within the impulse approximation, and with momentum assignments
corresponding to the quark triangle diagram of Fig.~\ref{fig1}, the vertex
function is given by the integral
\beqar
\Lambda _{\mu \nu }(P,Q)=-\mbox{tr}\int \frac{d^4k}{(2\pi )^4} &&
S(k-P-Q)\Gamma _{\mu }( k-\frac{P}{2}-\frac{Q}{2};-P-Q)
S(k)\nonumber \\
&&\times \Gamma _{\nu }( k-\frac{Q}{2};Q)S(k-Q)i\gamma _5
\tau _3\Gamma _{\pi }( k-\frac{P}{2}-Q;P) .\label{int}
\eeqar
Here $S(k)$ is the dressed quark propagator, and $\Gamma _{\pi }(p;q)$
and $\Gamma _{\mu }(p;q)$ represent respectively the pion-quark
Bethe-Salpeter amplitude and the dressed quark-photon vertex corresponding
to incoming boson momentum $q$ and relative $\bar{q}q$ momentum $p$.  The
trace is over spin, flavor ($u$ and $d$ only) and color.  We
require that the dynamical quantities $S$, $\Gamma _{\mu }$ and
$\Gamma _{\pi }$ be mutually consistent with electromagnetic gauge
invariance, dynamical chiral symmetry breaking, confinement and the known
behavior of perturbative QCD.

This generalized impulse approximation, with consistently dressed elements,
can be derived as the lowest-order term in the meson loop expansion of
the electromagnetically gauged version~\cite{FT} of the model-QCD
field theory known as the Global Color-symmetry Model (GCM).~\cite{cr85,gcm}
This replaces the gluon sector by a momentum dependent finite range
effective gluon propagator, thus
formalizing the Abelian approximation to QCD and properly embedding the chiral
anomalies \cite{gcm}.
For the quark propagator, written as
$S(p)=-i \gamma\cdot{p}~\sigma_{V}(p^2) + \sigma_{S}(p^2)$,
we employ the parameterized amplitudes~\cite{pionff}
\beqar
\bar{\sigma}_{S}(x)&=&c~e^{-2x}+\frac{1-e^{-b_{1}x}}{b_{1}x}
\frac{1-e^{-b_{3}x}}{b_{3}x} \Bigl( b_{0}+b_{2}
\frac{1-e^{-b_4 x}}{b_4 x}\Bigr)+\frac{\bar{m}}{x+\bar{m}^2}
\bigl(1-e^{-2(x+\bar{m}^2)}\bigr) \label{ss}	\\
\bar{\sigma}_{V}(x)&=&\frac{2(x+\bar{m}^2)-1+e^{-2(x+\bar{m}^2)}}
{2(x+\bar{m}^2)^2}-c~\bar{m}~e^{-2 x} \label{sv},
\eeqar
with $x=p^2/\lambda^2$, $\bar{\sigma}_{S}=\lambda \sigma_{S}$,
$\bar{\sigma}_{V}=\lambda^2 \sigma_{V}$, $\bar{m}=m/\lambda$, where
$m$ is the bare quark mass and $\lambda$ is the momentum scale.
The resulting $S(p)$ is an entire function in the complex momentum plane, a
sufficient condition for confinement, as it ensures the absence of quark
production thresholds in S-matrix amplitudes.~\cite{RWK}
When $b_0=b_2=0$, this
propagator accurately represents the entire function solution of the
simple, infra-red dominant, confining model SDE developed in Ref.\cite{BRW}.
The more general parameterization with $b_0,~b_2\neq 0$ allows a good
representation, at moderate momenta, of the behavior found in
realistic SDE studies~\cite{RW} while remaining consistent with the
requirements of perturbative QCD in the deep Euclidean region up to $\ln(p^2)$
corrections.
The parameters are $\lambda=0.516$~GeV, $c=0.0406$, $m=6.1$~MeV, and
$(b_0,b_1,b_2,b_3,b_4)=(0.118,2.51,0.525,0.169,1\times 10^{-4})$.
The fitted soft chiral physics quantities produced by this parameterization
are~\cite{pionff} $f_\pi=83.9$~MeV, ~$<\bar{q}q>_{1{\rm GeV}^2}=(211$~MeV$)^3$,
{}~$m_\pi=127$~MeV, ~$r_\pi^{em}=0.596$~fm together with reproduction of the
experimental $\pi \pi$ scattering lengths to within $20\%$.  In this
approach, choice of quark propagator parameters is equivalent to an implicit
choice of effective gluon propagator underlying the model field
theory.~\cite{RW}

With $S(p)^{-1}=i \gamma\cdot p~A(p^2)+B(p^2)+m$, representations for
$\Gamma_\mu$ and $\Gamma_\pi$ may be obtained in terms of $A$
and $B$ while satisfying the constraints of gauge invariance and
dynamical chiral symmetry breaking.  In particular, the chiral limit
($m=0$) result for the mass-shell $\Gamma_\pi$ is
\mbox{$\Gamma_\pi(p;P^2=0)=B(p^2,m=0)/f_\pi$} since the
SDE for $B(p^2)$ and the Bethe-Salpeter equation for $\Gamma_{\pi}$ become
identical.~\cite{DelScad}  For finite $m$ we use
\mbox{$\Gamma_\pi(p;P)\approx B(p^2,m)/f_\pi$}
since this produces the PCAC determination of $m_\pi$.
We use the Ball-Chiu~\cite{BC} ansatz for the dressed quark photon vertex
which is \mbox{$\Gamma_\mu(k;Q)=\hat{Q} \bar{\Gamma}_\mu(k;Q)$},
where \mbox{$\hat{Q}=\frac{1}{2}\bigl(\tau_3 + \frac{1}{3}\bigr)$} is the quark
charge operator, and
\beq
\bar{\Gamma}_{\mu}(k;Q)=-i\gamma_{\mu}\frac{1}{2}\Bigl(A(k_+)+A(k_-)\Bigr)
+ \frac{k_{\mu}}{k \cdot Q}\Bigl[ i\gamma \cdot k
\Bigl(A(k_-)-A(k_+)\Bigr) +
\Bigl(B(k_-)-B(k_+)\Bigr) \Bigr] \label{PQV}
\eeq
with \mbox{$k_\pm=k\pm \frac{Q}{2}$}.  This vertex satisfies the Ward-Takahashi
identity, transforms correctly and has the correct perturbative limit.
The above mutually consistent model for $S$, $\Gamma_\pi$ and
$\Gamma_\mu$ has been shown to provide an excellent description of the data
for the spacelike pion charge form factor at the level of the corresponding
impulse approximation.~\cite{pionff}

The $\pi^0 \gamma \gamma$ vertex function in (\ref{int}) is now completely
specified in terms of the quark propagator.  The integral is devoid of
spurious quark threshold singularities, is naturally convergent and is
evaluated by numerical quadrature.  At $Q^2=0$, our
numerical evaluation of the coupling constant yields $g_{\pi^0 \gamma \gamma}=
g(0,-m_\pi^2)=0.497$ in agreement with the previous application of this
model~\cite{pionff}, and in good agreement with the experimental value
$0.504 \pm 0.019$.  The chiral limit of this approach has been
shown~\cite{pionff} to correctly incorporate the exact
result $g(0,0)=1/2$ produced by the
axial anomaly independent of the form of quark propagator amplitudes.
Our numerical evaluation of the chiral limit reproduces this result.

\sect{3}{The Transition Form Factor.}
The obtained shape of the form factor $F(Q^2)=g(Q^2,-m_\pi^2)/g(0,-m_\pi^2)$
at the pion mass-shell is
displayed in Fig.~2 by the solid line along with the CELLO collaboration
data.~\cite{CELLO}  Also shown there is a recent result from a QCD sum rule
approach \cite{rady}, and a monopole form~\cite{bl2} that interpolates from the
leading
asymptotic behavior $8 \pi^2 f_\pi^2 /Q^2$ argued from the perturbative
QCD factorization approach.~\cite {bl1}  In the latter two approaches
there is considerable ambiguity due to: A) the unknown momentum scale at
which perturbative behavior should set in; and B) assumptions for the pion
wavefunction and how it should evolve to non-perturbative momentum
transfer.~\cite{rady}
Within the present approach, both the photon coupling and the produced
pion wavefunction evolve with $Q^2$ in a way determined by the evolution
of the dressed quark propagators.  This produces, in a single expression,
both the ultra-violet behavior required by perturbative QCD and the
infra-red limit dictated by the axial anomaly.
\begin{figure}
\unitlength1.cm
\begin{picture}(14,9)(-16.0,1.5)
\includegraphics{ggpifig2.ps}
\end{picture}
\caption{The $\gamma^* \pi^0 \gamma$ transition form factor at the pion
mass-shell.  The data is from Ref.~\protect\cite{CELLO}, the solid line is the
present work, the dashed line is the monopole shape which interpolates
in from the leading asymptotic behavior produced by perturbative QCD
factorization \protect\cite{bl2}, and the dot-dashed line is a recent QCD
sum rule calculation \protect\cite{rady}.}
\label{fig2}
\end{figure}

The predicted scale for transition from perturbative to non-perturbative
regimes can be unfolded as follows.  The asymptotic behavior produced by the
quark triangle diagram can be obtained in closed form without factorization
by application of Feynman integral techniques with bare photon coupling and
free quark propagators thereby preserving gauge invariance.
We note that the employed pion Bethe-Salpeter amplitude $B(p^2,m)/f_\pi$
has the correct leading power law behavior $m\lambda^2/p^2f_\pi$ which
implements the hard gluon contribution that dominates perturbative QCD.  The
asymptotic form
\mbox{$F(Q^2)\sim A~[Q^2+B+C~\ln(Q^2/Q_0^2)/Q^2]^{-1}$}, where
\mbox{$Q_0=$}\mbox{$1~{\rm GeV}^2$}, is found to incorporate the first three
leading terms obtained this way and also provides an excellent fit to
the numerical results for $Q^2\geq 10$~GeV$^2$ with $A=0.53$~GeV$^2$,~$B=3.1~
{\rm GeV}^2$ and $C=-3.8~{\rm GeV}^4$.  Our direct calculation of this
anomalous process; i.e. without employing a factorization prescription,
yields \mbox{$A/f_{\pi}^2 = 75$}, in agreement with the normalization
(\mbox{$8\pi^2= 79$}) and leading
asymptotic behavior ($Q^{-2}$) obtained from factorization.~\cite{bl1}
It must be emphasized, however, that the interplay of soft and hard mechanisms
prevents the $A/Q^2$ term from dominating to
better that $85$\% until \mbox{$Q^2 \geq 20~{\rm GeV}^2$}.  The recent QCD
sum rule approach~\cite{rady} assumed that at \mbox{$Q^2 = 3~{\rm GeV}^2$}
the leading term provides an accurate foundation for generating the
soft physics behavior.  This is not borne out by the present investigation;
the non-leading contribution is still providing $50\%$ of the strength at
that scale.

Off the pion-mass shell (\mbox{$P^2 > -m_{\pi}^2$}), we use for the pion vertex
\mbox{$\Gamma_\pi(p;P) \approx$}
\mbox{$ B(p^2,m)/\sqrt{Z(P^2)}$}, where $Z(P^2)$
is obtained from the calculated tree-level inverse propagator
$(P^2+m_{\pi}^2)Z(P^2)$ for the $\bar{q}q$ mode in the pion channel
produced by the model.~\cite{FT}   With this convention, the departure of $Z$
from its mass-shell value $f_\pi^2$ defines a vertex that communicates
with pions via the conventional point propagator to facilitate subsequent
applications.  The calculated \mbox{$\gamma^* \pi^0 \rightarrow \gamma$}
transition form factor $F(Q^2,P^2)=g(Q^2,P^2)/g(0,-m_\pi^2)$,
for a significant range of virtual pion momenta
($P^2\geq-m_\pi^2$), is well represented as a suppression of the
mass-shell form factor $F(Q^2)$ by the expression
\mbox{$F(Q^2,P^2)\simeq F(Q^2)~f(P^2)$}.  The calculated suppression factor
is shown in Fig.~3.  The form
\beq
f(P^2)=\frac{1}{1+(P^2+m_\pi^2)/a^2-(P^2+m_\pi^2)^2/b^4},
\label{f}
\eeq
with \mbox{$a= 1.38~{\rm GeV}$}, and \mbox{$b= 1.55~{\rm GeV}$}
provides a fit to within $7\%$ for
$-m_\pi^2\leq P^2\leq 1~{\rm GeV}^2$ and \mbox{$Q^2 \leq 2.5 {\rm GeV}^2$}.
This is more than sufficient to cover the range of minimum $P^2$ allowed by
typical electron scattering angles considered to favor the $\gamma^* \pi^0
\rightarrow \gamma$ mechanism in a feasibility study for an experiment on
virtual Compton scattering on a proton target at CEBAF.~\cite{cebaf}
This prediction of the dependence upon virtual pion momentum is, of course,
subject to uncertainities associated with the definition of an off-mass-shell
$\bar{q}q$ correlation in the pion channel.
An illustration of the uncertainties can be obtained by comparing these
results with those obtained in the point-pion limit
($\Gamma_{\pi}=$~constant).  We find that this leads to at most a $12$\%
decrease of the suppression factor for the momenta considered in Fig.~3.
\begin{figure}
\unitlength1.cm
\begin{picture}(14,9)(-16.0,1.5)
\includegraphics{ggpifig3.ps}
\end{picture}
\caption{The reduction factor $g(Q^2,P^2)/g(Q^2,-m_\pi^2)$ for the
$\gamma^* \pi^{0} \gamma$ transition produced by extrapolating the pion
off its mass-shell in the space-like direction.  For any $\gamma^*$ momenta
between those shown, the results lie between the two curves.}
\label{fig3}
\end{figure}

\sect{4}{Summary.}
We present here for the first time a calculation of the
$\gamma ^* \pi^{0}\rightarrow \gamma$ form factor that maps out the off-shell
behavior with virtual pion momentum.   We find that the
dependence on the virtual-pion momentum is smooth and well described by
a simple suppression factor, which is qualitatively independent of
the details of the pion interpolating field.

A previous quark triangle diagram study \cite{ibg} was limited
to the pion mass shell.  Further distinguishing
features of the present work are that nonperturbative dressing of the photon-
quark vertex is included, the quark propagators have confining dynamical
self-energy amplitudes rather than a constant constituent mass, and
the pion Bethe-Salpeter amplitude is completely determined by the quark
propagator as required by dynamical chiral symmetry breaking.
The correct mass-shell axial anomaly is naturally generated by our approach,
the $Q^2$ dependence is in reasonable accord with the available data,
and no parameters are adjusted to achieve this.
A significant result of our study is that for this anomalous process, soft
nonperturbative effects remain significant for \mbox{$Q^2< 20~{\rm GeV}^2$}.
For the charge form factor $F_\pi(Q^2)$, a similar scale is produced from
the same model.~\cite{pionff}

We conclude that the generalized impulse approximation captures the dominant
spacelike physics of both the elastic and transition electromagnetic form
factors of the pion.
With more reliable data for the transition form factor, adjustment of the
present quark propagator model parameters should allow a high quality fit
to both the transition and elastic pion form factors as well as the other
pion data mentioned.  Since this approach is encoded in a QCD-based model
field theory, applications to other hadronic processes not dominated by
chiral symmetry (e.g. vector meson masses and mixing~\cite{romix}) can proceed
with a minimum of new parameters.

\vspace*{0.5\baselineskip}\hspace*{-\parindent}{\bf
Acknowledgments}\hspace*{\parindent}
We acknowledge very helpful discussions with A. Afanasev.  This work was
supported in part by the National Science Foundation under
Grant Nos. PHY91-13117, INT92-15223 and PHY94-14291.  This work was
also supported by the US Department of Energy, Nuclear Physics Division, under
contract Nos. W-31-109-ENG-38, and DE-FG06-90ER40561.


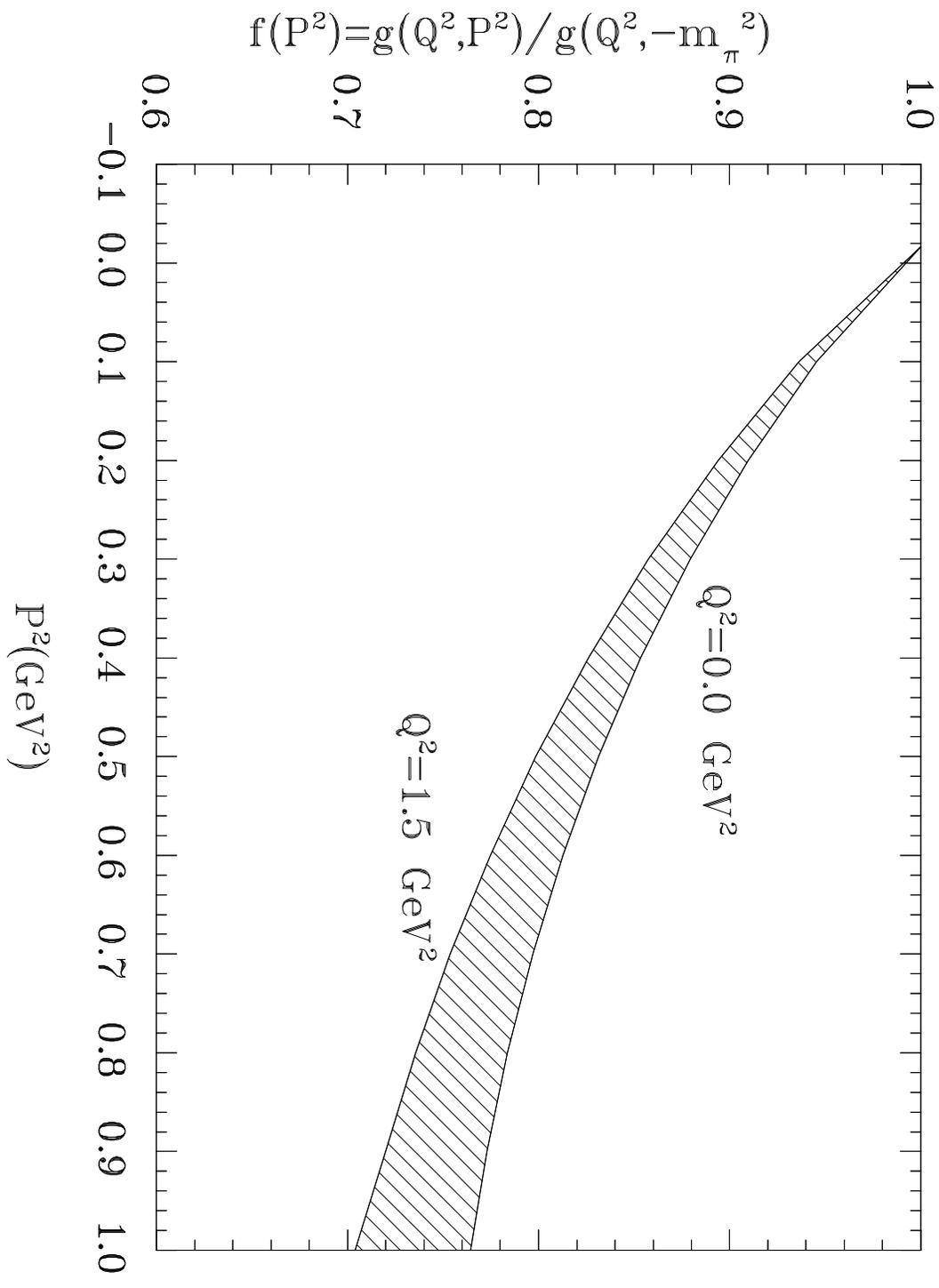
\end{document}